\documentclass[
    10pt,
    floatfix,
    twocolumn,
    superscriptaddress,
    groupedaddress,
    prl,
    a4paper,
    showpacs
]{revtex4}

\usepackage{graphics,graphicx}
\usepackage{amssymb,amsmath}
\usepackage{color}
\usepackage{bm}
\usepackage{float}
\usepackage{epsfig}
\usepackage[T1]{fontenc}
\usepackage{subfigure}
\newcommand{\w}{\omega}
\newcommand{\igr}{\includegraphics}

\newcommand{\etal}{\mbox{\textit{et al.}}}

\newcommand{\correlator}[1]{\langle#1\rangle}

\newcommand{\beq}{\begin{equation}}
\newcommand{\eeq}{\end{equation}}
\newcommand{\beqa}{\begin{eqnarray}}
\newcommand{\eeqa}{\end{eqnarray}}

\newcommand{\mev}{\textrm{meV}}
\newcommand{\muev}{\textrm{$\mu$eV}}

\newcommand{\M}{\textbf{M}}

\newcommand{\A}{\textbf{A}}

\newcommand{\figref}[1]{Fig.~\ref{#1}}

\begin{document}
\title{Localized edge vibrations and edge reconstruction by Joule heating in graphene nanostructures}



\author{M. Engelund}
\email{mads.engelund@nanotech.dtu.dk}
\affiliation{DTU Nanotech,
Department of Micro and Nanotechnology, Technical University of
Denmark, {\O}rsteds Plads, Bldg.~345E, DK-2800 Kongens Lyngby,
Denmark}

\author{J. A. F\"urst} \affiliation{DTU Nanotech,  Department of
Micro and Nanotechnology, Technical University of Denmark,
{\O}rsteds Plads, Bldg.~345E, DK-2800 Kongens Lyngby, Denmark}

\author{A.~P. Jauho}
\affiliation{DTU Nanotech, Department of Micro and Nanotechnology, Technical University
of Denmark, {\O}rsteds Plads, Bldg.~345E, DK-2800 Kongens Lyngby, Denmark}
\affiliation{Department of Applied Physics, Helsinki University of Technology, P.O.Box
1100, FI-02015 TKK, Finland}

\author{M. Brandbyge}
\affiliation{DTU Nanotech, Department of Micro and Nanotechnology,
Technical University of Denmark, {\O}rsteds Plads, Bldg.~345E,
DK-2800 Kongens Lyngby, Denmark}

\date{\today}
\begin{abstract}
Control of the edge topology of graphene nanostructures is critical to graphene-based electronics. A means of producing atomically smooth zigzag edges using electronic current has recently been demonstrated in experiments [Jia {\it et al.}, Science {\bf 323},
1701 (2009)]. We develop a microscopic theory for current-induced edge reconstruction using density functional theory. Our calculations provide evidence for localized vibrations at edge-interfaces involving unpassivated armchair edges. We demonstrate that these vibrations couple to the current, estimate their excitation by Joule heating, and argue that they are the likely cause of the reconstructions observed in the experiments.
\end{abstract}

\maketitle

Graphene, a single sheet of carbon atoms in a hexagonal lattice, is a material currently
under intense scrutiny \cite{Geim2007, Castro2009}. Graphene is interesting not only
because of its exotic material properties, but even more so due to its potential use in
future electronic components.  Graphene electronics has a tremendous potential
\cite{Avouris2007,Geim2009}, but its practical realization requires the ability to
manufacture graphene nanostructures in a controlled and efficient manner. The topology
of graphene edges plays a fundamental role in determining the electronic and transport properties of
these devices \cite{Nakada1996, Fujita1996, Wakabayashi2009}. Thus, the control and stability of edges is
crucial for further development of graphene-based electronic devices. Recent experiments
\cite{GiMeEr.2009,Warmer2009} show that simple armchair and especially zigzag edges are
the most commonly occurring edge structures, and that their formation and dynamics is
strongly influenced by the energetic electrons in a transmission electron microscope
(TEM). However, also intermediate reconstructed forms exist
\cite{KoMaHa.2008,KoMaHa.2009}. In an important recent experiment Jia {\it et al.}
\cite{Jia2009} demonstrated the formation of smooth zigzag edges from disordered edges
in graphene in the presence of an electronic current. The possibility of an {\it in
situ} fabrication process, as suggested by this experiment, is very attractive. However,
at the same time the devices should remain stable in the presence of electrical current
for reliable operation, further underlining the importance of understanding the
microscopic edge reconstruction mechanisms.


\begin{figure}[t!]
  \centering
  \igr[width=.95\columnwidth,clip]{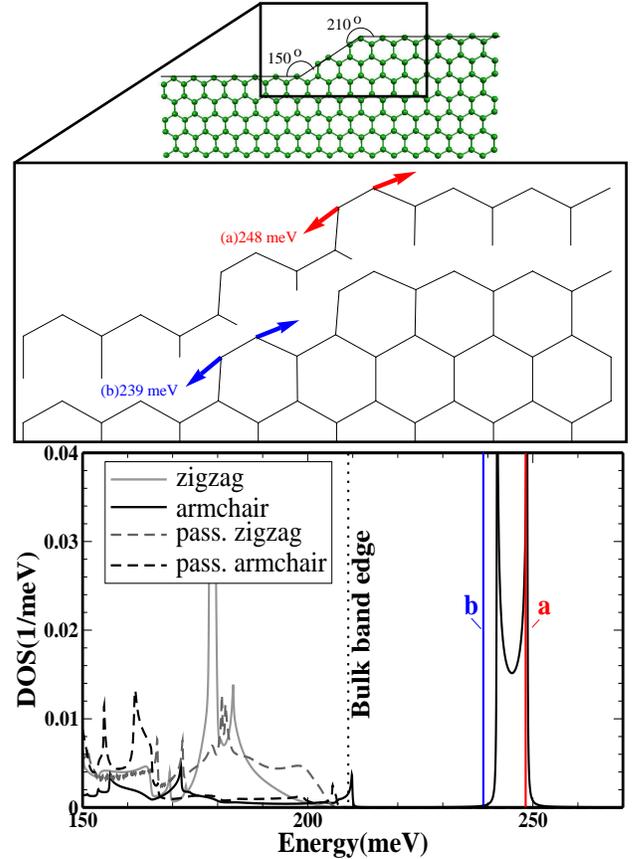}
  \caption{(Color online) [Top] The ZAZ structure.
  [Middle] The two modes lying outside the bulk bands, shown
  in red (a) and blue (b), respectively.
  Atomic displacements of less than $5\%$ of the total amplitude
  are not shown.
  [Bottom] The energies of the two localized modes
  compared with the local vibrational DOS of the outermost carbon atoms on
  infinite graphene edges. For H-passivated edges additional 1-D bands are
  found at \mbox{$\sim$380 meV} (not shown).}
  \label{phonon}
\end{figure}

\begin{figure}
  \centering
  \igr[width=.95\columnwidth]{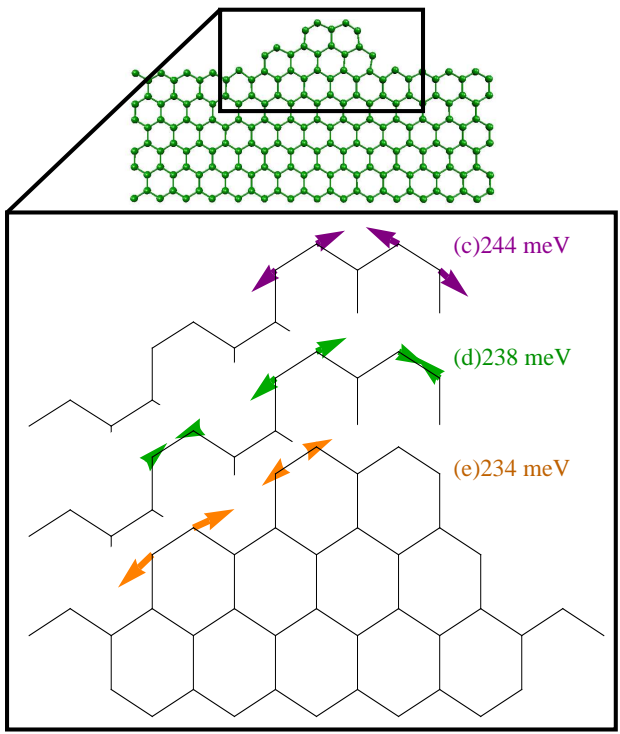}
  \caption{(Color online) [Top] The ZAZZZ structure.
  [Bottom] The three ADMs, shown
  in purple (c), green (d) and orange (e), respectively.
  Atomic displacements of less than $5\%$ of the total amplitude
  are not shown.}
  \label{phonon2}
\end{figure}

In this Letter we present an {\it ab inito} study of current-induced
edge reconstructions in systems, where armchair and zigzag edges are
adjacent. As the first example, consider the graphene nanoribbon
(GNR) junction shown in Fig.~\ref{phonon}. A zigzag-GNR to the left
is connected to a wider zigzag-GNR to the right by an armchair edge
(ZAZ-structure). Note that the edges are not passivated (we return
to this point below). The second example is a ZAZZZ-system (see
\figref{phonon2}) where an extra zigzag edge is inserted. These two
structures are chosen to mimic the experimental situation, and to
test the generality of the trends found in our calculations.

Using density functional theory \cite{Soler2002,Brandbyge2002} we
shall demonstrate that strong local Joule heating occurs in systems
of this kind for voltage biases and currents of the same order of
magnitude as in the experiment by Jia {\it et al.} \cite{Jia2009}.
In order to have significant Joule heating two conditions are of
importance. Firstly, localized vibrations are necessary in order to
spatially concentrate the energy. Secondly, the electronic subsystem
must couple strongly with the local vibrations and be locally out of
equilibrium in order to provide energy to the vibrations. As we
shall show, the structures shown in Figs.~\ref{phonon} and
\ref{phonon2} indeed exhibit localized vibrational modes. These
modes originate from "armchair-dimers", defined as \mbox{C-C} dimers
coordinated in the same way as the outermost atoms of a
non-passivated armchair edge.  In what follows, we call these modes
"armchair dimer modes" (ADMs).
Next, we show that zigzag-armchair junctions exhibit strong local
scattering of the current carrying electrons such that we can expect
a local voltage-drop across the junction. Finally, we estimate the
heating of the ADMs in the two model systems and argue that the
heating of the ADMs is the likely cause of the reconstruction of the
edges observed in the experiment by Jia \etal~\cite{Jia2009}.

Let us now address quantitatively the existence of localized
vibrational modes in systems with mixed zigzag and armchair edges.
Using finite displacement calculations \cite{finite_displ} we find
two such modes (see \figref{phonon}) for the ZAZ system with a
vibrational energy of $\sim$250~meV. These modes are strongly
localized to the outermost atoms of the armchair edge: already for
nearest neighbors the mode amplitude has dropped by $\sim$$85~\%$.
The modes are truncated versions of the vibrational edge states that
give rise to the quasi-1D band in the density of states (DOS)
\cite{vibrational_dos} of the infinite armchair edge (full black
curve in \figref{phonon}, see also Ref. \onlinecite{Zhou2007}).
These two ADMs are energetically localized since both the infinite
zigzag edges (full grey curve in Fig. \ref{phonon}) and bulk
graphene (band edge shown as a dotted line) have a vanishing DOS
above $\sim$200~meV and thus cannot cause energy broadening above
this energy in the harmonic approximation. The modes outside the
bulk band are only broadened by interactions with electrons and, at
high temperature, by anharmonic interactions.

For the ZAZZZ system we identify three localized modes
(Fig.~\ref{phonon2}) which again involve two-coordinated armchair
dimers. For this case the vibration amplitude is significant not
only on the armchair edges but also at the $240^\circ$ ZZ edge.

Hydrogen passivation is predicted to play a role for the edge
structure\cite{WaSeSa.2008}. However, we expect the ADMs to be localized even if they
are adjacent to hydrogen-passivated edges. We find that the passivated zig-zag and
armchair edges do {\it not} have vibrations with energies $\sim$250~meV in their DOS
(dashed black and grey curves in \figref{phonon}, respectively), in agreement with the
empirical-potential calculation in Ref. \cite{Vandescuren2008}, and thus cannot provide
harmonic damping. We constrain our studies to non-passivated edge interfaces since there
were no signs of hydrogen playing a role in the experiments by Jia {\it et al.}
\cite{Jia2009} and we also expect hydrogen to desorb at much lower temperatures than
where the breaking of carbon-carbon bonds takes place. The barrier for hydrogen
desorption from an armchair edge is lower compared to that of the outermost \mbox{C-C}
unit, $4$~eV (Ref.~\cite{Dino2004}) and $6.7$~eV (Ref.~\cite{Jia2009}), respectively.

\begin{figure}
\includegraphics[width=0.95\columnwidth,viewport=50 30 490 280,clip,angle=0]{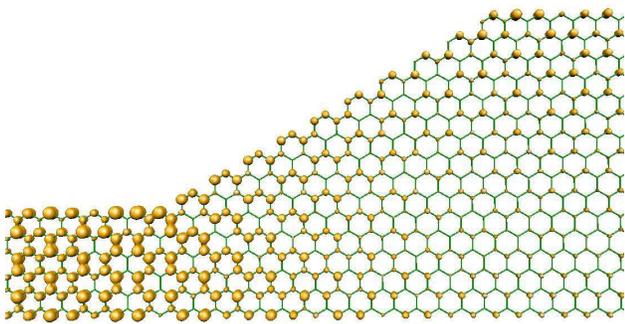}
\caption{\label{electron}(Color online) The absolute square of the minority spin
  left-to-right scattering states at the Fermi level for a large ZAZ system,
  corresponding to a transmission of 0.15.}
\end{figure}

In addition to the existence of localized vibrational modes, the
occurrence of local Joule heating also requires local scattering of
electrons. By NEGF-DFT transport calculations
\cite{Brandbyge2002,Haug2008,spin_pol}, we find that the scattering
states localized at the zigzag edge are interrupted at the
zigzag-armchair junction, illustrated by the square modulus of the
transmitting electronic scattering states (eigenchannels
\cite{Paulsson2007}) at the Fermi level, shown for the minority spin
in Fig.~\ref{electron}. Ref. \cite{Wakabayashi2001} shows, by
extensive tight-binding calculations, that scattering can in general
be expected at armchair-zigzag edge interfaces and that scattering
increases with the length of the armchair edge.  Since the same
behavior is predicted for different systems with two different
methods ({\it ab initio} vs. tight-binding), it seems that
back-scattering at armchair-zigzag interface indeed is a generic
feature of these systems.

\begin{figure}
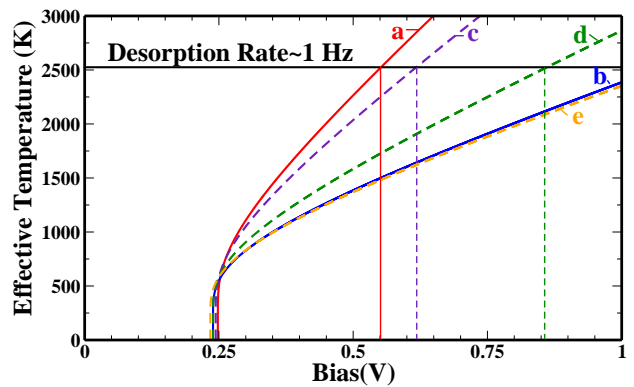

  \centering
  \igr[width=.95\columnwidth,clip]{./temperature}
  \caption{(Color online) The effective temperature as a function
  of bias for the localized modes for the ZAZ (full red and blue curves)
  and the ZAZZZ system (dashed purple, green and orange curves). The full black
  horizontal line indicates the uniform temperature where the decay rate of the
  outer \mbox{C-C} dimer of an armchair reaches 1 Hz.}
  \label{temperature}
\end{figure}

We next estimate the heating of the ADMs in the two model systems.
If the anharmonic couplings are neglected, the mean steady-state
occupation, \mbox{$\correlator{n^\lambda}(V)$}, can be calculated
from the ratio of the current-induced phonon emission rate,
$\gamma^\lambda_{\rm em}(V)$, and the effective phonon damping rate,
$\gamma^\lambda_{\rm damp}(V)$ (here $V$ is the voltage drop across
the scattering region). Since the investigated modes lie outside the
vibrational bulk band, the damping due the bulk phonon reservoir
vanishes. Assuming zero electronic temperature and
energy-independent scattering states within the bias window, the
emission rate is \cite{Frederiksen2007a,energy_indep}
\begin{equation}
\gamma^\lambda_{\rm
em}(V)=\frac{eV-\hbar\w_\lambda}{\hbar\pi}\theta(eV-\hbar\w_\lambda)
{\rm Tr}[\M^\lambda \A_L \M^\lambda \A_R ]\,.
\end{equation}
Here, $\hbar\w_\lambda$ is the vibrational energy of mode $\lambda$,
$\M^\lambda$ is the coupling of the mode to the electronic degrees
of freedom calculated by finite difference techniques, and
$\A_{L/R}$ is the electronic spectral density of left/right moving
electrons, evaluated at the Fermi level by NEGF-DFT transport
calculations \cite{Brandbyge2002,Haug2008,spectral_dens}. The rate
$\gamma^\lambda_{\rm damp}(V)$ is \cite{Frederiksen2007a}
\begin{equation}
\gamma^\lambda_{\rm damp}(V)=\frac{\w_\lambda}{\pi}{\rm
Tr}[\M^\lambda \A \M^\lambda \A]\,,
\end{equation}
where $\A=\A_L+\A_R$ is the total electronic spectral density at the
Fermi level. The occupation $\correlator{n^\lambda}(V)$ becomes
\begin{equation}
\correlator{n^\lambda}(V)=\frac{1}{2}\theta(eV-\hbar\w_\lambda)(\frac{eV}{\hbar\w_\lambda}-1)s^\lambda\,,
\end{equation}
where $s^\lambda=\frac{2{\rm Tr}[\M^\lambda \A_R \M^\lambda \A_L]}
 {{\rm Tr}[\M^\lambda \A\M^\lambda \A]}$
is a dimensionless heating parameter that can vary from 0 (no
heating) to 1 (maximal heating). By assuming that $n^\lambda(V)$ is
Bose distributed, one can extract an effective temperature of the
mode, $T^\lambda_{\rm eff}(V)$, which we plot in
\figref{temperature} for the localized modes identified in
\figref{phonon} and \figref{phonon2}. The calculated electronic
damping rates and heating parameters are shown in
Table~\ref{tab:charateristics}.


The mode temperatures should be compared to the uniform temperature,
$T_d$, needed to destabilize the system on a time-scale relevant to
the experimental conditions, which we judge to be of the order of
seconds. Thus we consider a corresponding rate of desorption of
carbon dimers, \mbox{$q\sim 1$~Hz}. We estimate  $T_d\sim$2500~K
(see \figref{temperature}) using the Arrhenius equation, $q=\nu
\exp[-E_a/(k_B T_d)]$, a characteristic attempt frequency $h\nu
=100~\mev$ (a typical phonon frequency), and an activation energy
\mbox{$E_a=6.7$~eV} \cite{Jia2009}.


\begin{table}
  \centering
    \begin{tabular}{| c | c | c | c | }
    \hline
    Mode & $\hbar\w_\lambda(\mev)$ & $\hbar\gamma_{\rm damp}^{\lambda}(\muev)$& $s^\lambda$\\
    \hline
    ZAZ, a&  248&99&0.77\\
    \hline
    ZAZ, b& 239& 63& 0.29\\
    \hline
    ZAZZZ, c& 244 & 19&0.63\\
    \hline
    ZAZZZ, d& 238 & 26 &0.38\\
    \hline
    ZAZZZ, e& 234 & 64 &0.28\\
    \hline
  \end{tabular}
  \caption{Mode characteristics.}
  \label{tab:charateristics}
\end{table}

We can compare the heating of the localized ADM to the modes
inside the bulk band by calculating their harmonic damping due to
their coupling to the bulk phonons \cite{Engelund2009}, and adding
this to the damping by the electrons. We find that this damping is
$1-100$ times the damping due to the electronic couplings leading to
temperatures typically below 1000 K even for a bias of 1 V. Since
this temperature yields desorption rates much lower than those seen
in the experiments we conclude that the Joule heating of the
harmonically damped modes cannot account for the reconstruction.
Instead, as \figref{temperature} shows, the ADMs reach a high
temperature at much lower biases, and can thus provide a channel for
local desorption. Furthermore, these modes are also more likely to
be involved in the desorption since they directly involve the
desorbing dimers.

A quantitative comparison with experiments would require a much more
sophisticated theory, beyond the scope of this Letter.  For example,
one should consider the anharmonic coupling of ADMs, and evaluate
the nonequilibrium electronic distribution function, from which the
actual potential profile in the sample can be extracted.



Finally, let us investigate the role of different types of edge
interfaces. Fig.~\ref{phonon} shows two examples: a $150^\circ$
zigzag-armchair interface, and a $210^\circ$ armchair-zigzag
interface. Intuitively one would expect the 2-coordinated dimer
directly at the $210^\circ$-corner to make this interface especially
prone to reconstructions. This is confirmed by our calculations of
their heating: we find that the $210^\circ$ modes (modes (a),(c) and
(d)) exhibit markedly stronger heating than modes associated with
the $150^\circ$ corner, the reason being their stronger coupling to
the current carrying electrons ($s^\lambda$ closer to 1 for these
modes). This scenario is consistent with two experimental
observations by Jia \etal \cite{Jia2009}. Firstly, they observe that
a $150^\circ$ interface survives even after massive reconstruction
has occurred (Ref.~\cite{Jia2009}, Fig. 2(D)). Secondly, certain armchair edges evaporate while others grow
longer. The theory outlined here predicts that the evaporating
armchair edges are the ones bordered by at least one $210^\circ$
junction. Conversely, zigzag edges and armchair edges bordered by
$150^\circ$ junctions would be the stable edges, and would grow as
the unstable armchair edges evaporate.

In conclusion, we have demonstrated how specific \mbox{C-C} dimers
can play a fundamental role in current-induced reconstruction in
graphene systems with mixed edges.
We show that these dimers give rise to spatially and
energetically localized modes, which give a natural explanation for
the low onset bias for reconstruction observed in the experiments
\cite{Jia2009}. Identifying the modes that heat up also allows us to
make predictions of the overall behavior of a graphene sample under
the influence of a current. Specifically, we predict that
zigzag-armchair junctions with an angle of $150^\circ$ would be more
stable than the junctions with a $210^\circ$ angle. We believe
reasoning along these lines could contribute towards a quantitative
understanding of other intriguing  edge structures, e.g. the
zigzag-reczag discussed recently in Ref. \cite{KoMaHa.2009}.

{\it Acknowledgements}.  We acknowledge useful correspondence with
the authors of Ref. \cite{Jia2009}.  APJ is grateful to the FiDiPro
program of the Finnish Academy. Computational resources were provided by the Danish Center
for Scientific Computing (DCSC).

\bibliography{graphene_vibrational}

\begin{thebibliography}{28}
\bibitem[{spi()}]{spin_pol}
\bibinfo{note}{Spin-polarized calculations with single-$\zeta$ basis and 200~Ry
  real-space mesh were used.}
\bibitem[{ene()}]{energy_indep}
\bibinfo{note}{The assumption of energy-independent scattering states is
  justified by observing that the transmission is approximately constant from
  -0.5~V to 0.5~V for the two systems.}
\bibitem[{spe()}]{spectral_dens}
\bibinfo{note}{The electronic spectral density was calculated with
  spin-polarized DFT.}
\end{thebibliography}

\clearpage
\end{document}